# Thousands of Positive Reviews: Distributed Mentoring in Online Fan Communities

**Julie Ann Campbell**
University of Washington
Seattle, USA
juliemu@uw.edu

**Cecilia Aragon**
University of Washington
Seattle, USA
aragon@uw.edu

**Katie Davis**
University of Washington
Seattle, USA
kdavis78@uw.edu

**Sarah Evans**
University of Washington
Seattle, USA
sarahe@uw.edu

**Abigail Evans**
University of Washington
Seattle, USA
abievans@uw.edu

**David P. Randall**
University of Washington
Seattle, USA
dpr47@uw.edu

## ABSTRACT
Young people worldwide are participating in ever-increasing numbers in online fan communities. Far from mere shallow repositories of pop culture, these sites are accumulating significant evidence that sophisticated informal learning is taking place online in novel and unexpected ways. In order to understand and analyze in more detail how learning might be occurring, we conducted an in-depth nine-month ethnographic investigation of online fanfiction communities, including participant observation and fanfiction author interviews. Our observations led to the development of a theory we term *distributed mentoring*, which we present in detail in this paper. Distributed mentoring exemplifies one instance of how networked technology affords new extensions of behaviors that were previously bounded by time and space. Distributed mentoring holds potential for application beyond the spontaneous mentoring observed in this investigation and may help students receive diverse, thoughtful feedback in formal learning environments as well.

**ACM Classification Keywords**
H.5.3. Group and organization interfaces: Theory and models.

## INTRODUCTION
In a 45-minute period during the afternoon on Monday, May 4, 2015, 100 new fanfiction stories were published on the online repository Fanfiction.net. Fanfiction is immensely popular among adolescents and young adults, and this popularity is evident in the millions of stories housed in online fanfiction repositories. Fanfiction.net is the 440th most popular website in the United States, and its visitors are overwhelmingly teens and emerging adults [1]. Fanfiction authors have written stories housed on Fanfiction.net that are several hundred thousand words in length, even longer than the original works on which the stories were based.

This abundance of literary production stands in sharp contrast to popular wisdom that today's young people are not interested in or even capable of long-form writing and use technology primarily for shallow interactions on social media [11,23,59,62]. What is it about fanfiction experiences that foster such engaged participation among youth? We set out to investigate why members of a generation often maligned for placing video games ahead of books are so engaged in this form of writing. During this analysis, we communicated directly with young fanfiction authors in addition to observing interactions between authors and readers over the course of nine months. In the process, we discovered evidence of a type of community-based learning and mentoring that is uniquely facilitated by the technological affordances of the Web. Moving beyond formal mentoring, people have developed a network-enabled form of giving and receiving feedback, actively correlating knowledge gleaned in small pieces from a large community to inform their fanfiction writing. As this knowledge is embodied in artifacts of interaction between community members, we draw on and extend Hutchins' framework of distributed cognition [30,31] and Dawson's mentoring framework [18] in naming this process *distributed mentoring*.

While "distributed mentoring" has been used previously by other scholars [16,40,50] to describe mentors who are geographically distributed or collectively based, our use of the term is novel because of its grounding in Hutchins' concept of distributed cognition, and because it is not just the people but also the mentoring itself that is distributed. We identify seven key attributes of distributed mentoring that distinguish it from traditional, offline forms of mentoring and tie it both to Hutchins' theory of distributed cognition and to the particular affordances of networked publics. These attributes are: *aggregation, accretion, acceleration, abundance, availability, asynchronicity,* and *affect*.

Grounded in a sociocultural tradition of theory and research, our work builds upon and extends previous research addressing the role of mentoring in affinity spaces and communities of practice [16,24,25,40,42,50] by delving into the specific processes by which knowledge is shared, the distinct affordances of networked platforms that facilitate this sharing, and how writers experience these processes of knowledge sharing. In this way, we extend prior work by articulating a new theoretical understanding of networked mentoring processes and how they differ from a traditional model of mentoring. We believe that the theoretical construct of distributed mentoring may be of benefit in understanding learning processes in other formal and informal learning environments.

## PREVIOUS WORK AND BACKGROUND
### Fanfiction
Scholars have explored fanfiction from a number of angles, including legality [47,55,57], literacy [8,35], instructional potential [12,44], sexuality [34,58], and identity [3,8]. Jenkins' [34] seminal investigation of fanfiction occurred more than 20 years ago as part of a larger study of fan culture. Challenging negative stereotypes of fans, Jenkins depicted them "as active producers and manipulators of meanings" [34]. He identified fanfiction as a prime example of how fans engage actively with original creative works by referencing, manipulating, and extending them in new and unexpected ways.

Fanfiction has existed since the 1960s, developing within the fandoms for television shows like Star Trek [34]. However, the distribution of fanfiction has substantially changed with the advent of internet-based fanfiction repositories. Removing the physical constraints of printing has lowered the barriers to reader critique, with the result that there may be hundreds or thousands of reviews submitted for a single fanfiction story. Additionally, the public, persistent, and immediate nature of online reviews allows for instant and highly interactive communication among reviewers as well as the fanfiction authors themselves. This dramatic decrease in the cost of providing feedback to authors has led to far greater volumes of information being exchanged among larger numbers of participants in online fanfiction communities, enabling a complex web of support, commentary, communication, and mentoring that could not have been possible with older technologies: what we term distributed mentoring.

Following Jenkins' investigation of fanfiction, Black [6–8] investigated language acquisition and identity formation among English language learners participating in online fanfiction communities. Black analyzed the dialogue between authors and readers in the form of story reviews and authors' notes. She found that these interactions helped the authors shape their learning space and identity. Black's focus, however, was primarily on how reviews affected individual authors' identity formation over time. We still lack insight into the complex structure of the network of fanfiction reviews and the impact that an entire community of reviewers may have on authors' creative processes and skill development. This topic is worth further exploration in light of fanfiction's popularity and the encouragement that authors profess to feel because of reader reviews. Understanding this phenomenon in more depth may provide insights into how young people could be encouraged to develop their skills as writers as well as their enjoyment of the writing process in arenas outside the fanfiction world. In this way, our work contributes to existing work exploring the use of fanfiction and other forms of fan creation to support learning in formal and informal educational settings [43,49].

### Affinity Spaces, Participatory Culture, and Connected Learning
The present study takes as its starting point the recognition that fanfiction communities function as affinity spaces [8,36]. Gee [24] described an affinity space as an "interest-driven passion-fuelled site," where personal differences in age, gender, or class are less important than the shared interest being explored. Like communities of practice [42], affinity spaces are distinct contexts of learning and identity formation where novices become experts through guided participation in the social structure of the community [24,25]. However, Black [8] distinguishes between the two by observing that in affinity spaces there are a range of different novice and expert roles that vary based on context, whereas in communities of practice, apprentices are being trained toward one type of full participation using a well-defined body of knowledge. In a fanfiction community, authors may be apprentices and masters at the same time, possessing expertise in one aspect of the craft, such as characterization, grammar, or canon knowledge, but requiring assistance in other aspects. In the present work, we expand on Gee's concept of affinity spaces and related work around networked practices of feedback [8,38,41,54] by exploring how the distinct affordances of networked technologies shape and transform mentoring processes in affinity spaces. In so doing, we demonstrate how the concept of distributed mentoring offers new insights into collaborative mentoring processes in networked environments.

The present work is also informed by participatory culture and connected learning principles, which networked affinity spaces are well positioned to support. According to Jenkins [36], a participatory culture is "a culture with relatively low barriers to artistic expression and civic engagement, strong support for creating and sharing creations, and some type of informal mentorship whereby experienced participants pass along knowledge to novices" [36]. Fanfiction communities promote participatory culture by virtue of their low barrier to entry—anyone of any age can write and publish a story—and their high level of peer support and mentoring. The concept of distributed mentoring extends Jenkins' work around participatory culture by delving more deeply into the relationship between the affordances of networked technologies and the mentoring processes we observed in fanfiction communities. In so doing, we draw on Ito et al.'s model of connected learning [33], which is defined by three core properties: learners come together around a shared purpose, a focus on production, and openly networked infrastructures. Our work illustrates how these three properties support and give rise to new forms of openly networked, collaborative mentoring.

**Mentoring**

Mentoring relationships represent a common thread in descriptions of affinity spaces, communities of practice, participatory culture, and connected learning. Mentors provide community participants with support in developing the skills and competencies necessary for full and productive participation in the community. In fanfiction communities specifically, informal peer mentors help authors improve their writing skills by providing feedback and answering questions [8].

The term "mentoring" has been used in so many different ways and contexts that experiences referred to as mentoring may be quite distinct from each other. As a result, Dawson [18] argued for the specification of mentoring models and provided a common framework for researchers to use in crafting their definitions. Mentoring is not limited to formal relationships between an older, more experienced mentor and a younger, less experienced mentee, which Dawson refers to as "traditional," but instead may take place between participants of all ages, with different levels of engagement and formality. Based on a survey of mentoring literature and feedback from mentoring professionals, Dawson's framework contains sixteen dimensions along which to classify mentoring relationships; objectives, roles, cardinality, tie strength, relative seniority, time, selection, matching, activities, resources and tools, role of technology, training, rewards, policy, monitoring, and termination. Drawing on Dawson's [18] mentoring framework, we classify fanfiction sites as informal, weakly tied, peer-based, many-to-one or many-to-many, CMC-only [21] mentoring intended to improve certain skills—in the present analysis, writing skills. The mentoring is informal and weakly tied because the engagement between authors may be spontaneous and limited to a single incident, or it may build into a longer relationship if the authors are interested. The community participants taking part in this mentoring are all peers in the same affinity space in which there is no prescribed formal hierarchy. The large number of participants involved in the mentoring indicates a many-to-one or many-to-many experience that is distinct from simply having multiple mentors.

The advent of networked technologies has opened up new possibilities for learning and mentoring that stretch Dawson's [18] framework and reveal its limitations. Stahl, Koschmann, and Suthers [53] referred to "fundamentally social technologies" to describe technologies with specific affordances that support social interactions and group learning [53]. With respect to mentoring in particular, affordances such as asynchronous communication, anonymity, and persistent, searchable content introduce advantages of computer-mediated mentoring versus traditional mentoring, including a record of interactions, equalization of status, and reduced interaction costs [21]. Communication within online fanfiction communities is by its very nature computer-mediated and, therefore, generates a persistent record of the interactions between authors and readers for both to learn from at their leisure. There is no overt distinction among authors on their profile pages; all authors may choose whether or not to include demographic information. In such a context, the age-based distinctions typical of offline settings are not as overt in online fanfiction communities, providing teens and emerging adults with unique opportunities to assume mentoring roles [32]. In addition to the decreased cost of reader interactions mentioned earlier, these qualities distinguish the mentoring occurring in fanfiction communities from more traditional, non-computer-mediated mentoring.

**Distributed Cognition and Mentoring**

Hutchins' [30] theory of distributed cognition is useful for understanding the dynamics involved in coordinated group processes, such as those that are found in affinity spaces like online fanfiction communities. Instead of considering the individual mind as the unit of cognitive analysis, distributed cognition takes into account a system consisting of the individuals as well as their "information environment," which includes the tools and artifacts used as part of cognition [31].

Since Hutchins' initial exploration in offline environments, distributed cognition has been applied to computer-mediated interactions [15] and promoted as a theoretical foundation for human-computer interaction [28]. In addition, Aragon and Williams [4] drew upon distributed cognition in their study of distributed affect in collaborative creativity, expanding the focus of communication to include not only cognition but emotion.

The durable and distributed nature of online text-based communication allows for easy discussion, coordination, and error checking between participants dispersed across the globe [19,53]. Distributed cognition aids the understanding of a specific type of interaction in fanfiction communities: mentoring among participants. A fanfiction author residing in a small town in the United States can receive reviews and advice from individuals in any location, at any time, and much like in a cognitive system, each individual in this mentoring system can build on the other participants' insights. Interactions among geographically dispersed participants are facilitated by their shared knowledge of and interest in a particular fandom, including its characters, plot points, and universe. Although this mentoring system features key attributes of distributed cognition, such as the embodied representation of information in artifacts of the system (e.g. reviews) and the existence of a cognitive ecosystem, distributed mentoring also includes many additional features enabled by the affordances of networked publics, such as the ability to easily set up asynchronous communication, or the aggregation of pieces of information that ordinarily might be too small to provide meaningful mentoring, but that taken as a whole enable learning and skill development.

The current study builds on previous work by investigating the distinct mentoring processes involved in online fanfiction communities. We focused our research on three popular fandoms: Harry Potter, Doctor Who, and My Little Pony: Friendship is Magic. Previous research related to these fandoms has focused on fan identities [39], gender and sexuality [26,58], civic engagement and fan activism [27,37], copyright issues [14,51], and fan community engagement [60], rather than on mentoring processes related to fanfiction writing in these fandoms.

## METHODS

Drawing on a rich set of data sources, we describe the learning and mentoring that take place in fanfiction communities from the perspective of the authors, and present the results of our analyses of fanfiction user groups and story reviews. We define learning as a process of deepening participation in a social practice, and mentoring as positive support for this process [33,36,42]. We detail the theory of distributed mentoring that grew out of our ethnographic investigation of fanfiction communities and suggest its potential impact for computer-mediated learning.

The following research questions guided this study:

1. *How and what do young people learn through their involvement in fanfiction communities?*
2. *What role does mentoring play in participants' learning, and what does this mentoring look like?*

Our investigation centered on interviews with a sample of 28 young fanfiction authors and a nine-month period of participant observations in fanfiction communities.

We selected our three fandoms based upon the diversity of each fandom's genre (e.g. science fiction, fantasy, and animation), medium (e.g. TV, book) and length of time in existence. Our selection was also guided by the research team's personal familiarity with the fandoms, as significant fandom knowledge was necessary for understanding the discussions and stories in these communities. We selected three fandoms instead of focusing on one to make sure that what we witnessed was not an aberration within a single fandom. Additionally, we chose widely popular fandoms to ensure the size of the community and amount of material was sufficient for our study. We purposely opted for fandoms based on original story material (versus music group or hobby-based fandoms) because they inspire the majority of fanfiction writing as readers seek to extend the story and utilize the original text as mentor text [20,35].

### Study Sites

Fanfiction.net is the largest fanfiction repository on the internet with over six million registered users; it hosts more than five million fanfictions across thousands of fandoms. As of July 2015, the largest fandom on Fanfiction.net was Harry Potter, with 719,000 fanfictions, followed by Naruto, with 386,000. In addition to hosting fanfiction, Fanfiction.net has social features, such as communities, forums, beta-reader listings, and a private messaging system. Users on Fanfiction.net have profile pages that list the stories they have authored, the stories and authors they have marked as favorites, and introductory profile messages.

FIMFiction.net is a single-fandom fanfiction repository dedicated to the television show My Little Pony: Friendship is Magic. As of July 2015, FIMFiction.net had 185,014 users and 86,009 published stories [22]. Once users create an account on FIMFiction.net, they can access user groups, which include a common fanfiction repository and a discussion forum. Any member can create a group and there are more than 7,000 groups in existence. Group topics range from support groups for writers with learning disorders to groups dedicated to favorite minor characters. FIMFiction.net user pages are more customizable than profile pages on Fanfiction.net, and users have the option to write blog entries, leave public comments on other users' pages, and create custom content modules and image galleries. Like Fanfiction.net, FIMFiction.net also has a private messaging system where users can communicate directly.

We also interacted with and observed the Doctor Who fanfiction repository *A Teaspoon and an Open Mind* (Whofic.com), the My Little Pony: Friendship is Magic discussion board Ponychan.net, Livejournal.com fanfiction communities, and Reddit.com fanfiction subreddits.

### Interviews

We conducted in-depth interviews with 28 fanfiction authors via the private messaging systems on Fanfiction.net and FIMFiction.net. Sixteen of the authors were members of Fanfiction.net and 12 were members of FIMFiction.net. Twelve authors wrote My Little Pony fanfiction, nine authors wrote Doctor Who fiction, and seven authors wrote Harry Potter fiction.

Following previous research involving online communities [17,45], we selected authors through a process of purposive sampling, whereby members of the research team identified individuals who were active authors and reviewers in one of the chosen fandoms. We did this by looking at published stories on Fanfiction.com and FIMfiction.com with at least a small number of reviews and then viewing the author's user profile to ascertain if they were currently active on the site, had regularly received reviews on their writing and were likely within the age and language demographic of the study.

We chose to target authors between the ages of 18-25 so that we might better identify individuals who had recently encountered—or were currently engaged in—a mentoring experience in their formative years; however, some authors did fall out of this range, with an average age of 22.8, the youngest 13 and the eldest 30. We also chose to interview authors who were native English speakers and who had been writing fanfiction for more than one year.

In order to protect the privacy of our subjects, and given the complex issue of gender binaries and fluidity—which is often the subject of fanfiction stories—we did not require them to identify their gender. We instead chose to seek a gender balance through the selection of our primary research sites: Fanfiction.net and FIMFiction.net. Fanfiction.net has a primarily female audience [1] while FIMFiction.net's audience is predominantly male [2]. Of those subjects who did openly offer a gender identity, 9 identified as female (all were from Fanfiction.net) and 3 identified as male (with 2 from FIMFiction.net and one from Fanfiction.net).

The interviews included questions about community participation, mentoring, writing skills, and identity. Adopting Charmaz's constructivist grounded theory approach that builds upon Glaser, Strauss, and Corbin's methods [13], and guided by our research questions [48,56], members of our research team created a list of dominant themes that emerged directly from a close reading of the interview transcripts [56]. Through weekly research team meetings in which we reviewed the transcripts, discussed relationships among the themes, and wrote memos reflecting upon our process and observations, we collectively distilled the themes into the following four categories: 1) learning from writing fanfiction; 2) fanfiction's impact on life; 3) mentoring; and 4) the importance of fanfiction reviews. We used these themes to guide the focus of our participant observations.

**Participant Observation**

We used participant observations to explore in situ the themes that surfaced in the interviews and to further refine our theory. Following Boellstorff et al. [9], we were transparent about our position as researchers throughout the process.

To start our participant observation and help integrate ourselves into the community [6,9], each team member wrote a short fanfiction on a fandom of his or her choice and published it on a relevant fanfiction repository. This exercise provided us with an understanding of the author's perspective and lent us credibility as fans when contacting other authors.

The member-created groups on FIMFiction.net served as a primary site for our participant observations. The five FIMFiction.net groups observed each had between 1,000 and 5,000 members and focused on writing help and discussion. Three groups were spaces for struggling authors to ask questions, one group was a "school" to help writers learn their craft, and the largest, most active group was a general writing community that allowed questions and discussion.

We observed these groups over a period of nine months, taking notes and screen captures of what we saw. Each team member spent 5-10 hours per week observing the communities. This time was spent reading stories and reviews of recently published works and older stories, viewing Reddit and community posts, and participating in writing events within the communities such as National Novel Writing Month (NaNoWriMo). Additionally, we communicated informally via private message with members of the FIMFiction.net groups to ask them questions about their participation; these members were chosen when interesting behavior or topics were observed. None of the individuals contacted informally through participant observation were also part of the interview phase of the study.

Our observations also included author and reviewer communications on Fanfiction.net. Because Fanfiction.net lacks the group activity of FIMFiction.net, reader reviews and author's notes are potentially the richest source of publicly viewable mentoring data on the site. Any reader can leave a review of a fanfiction on Fanfiction.net, either anonymously without creating an account or through a registered account. It is common practice for authors to leave notes either at the beginning or end of their work, and these notes will sometimes reference comments they have received in previous reviews or ask for reviewers' input on specific aspects of the work. We spent 10-20 collective hours each week reading stories, reviews, profiles, and author's notes on Fanfiction.net.

Our analyses of the participant observations served two primary purposes: 1) to corroborate themes surfaced in the interviews and 2) to examine processes of interpersonal communication among members of the fan communities. During the nine-month period of data collection, our team conducted a cumulative total of over a thousand hours of participant observation, generating several hundred pages of field notes, memos, and digests. This documentation was shared amongst the team and discussed in weekly team meetings.

During these discussions, we identified and highlighted evidence related to the dominant themes surfaced in the interviews. We paid particular attention to processes of mentoring among community participants. During the course of these weekly discussions and follow up analytic memos, we noted that mentoring appeared to be "distributed" across participants. Authors received guidance from multiple reviewers, and these reviewers sometimes referenced and built on comments left by earlier reviewers. Through this analytic process [13,48], we distilled seven attributes that together account for the full range of mentoring processes that we observed.

In the following sections, we report the dominant themes that emerged from our analysis of the author interviews, followed by a description of the seven attributes that we identified through discussions and analytic memos following our participant observations. All quotations included are unmodified, unless obfuscation was necessary for anonymity, and as a result may contain typographical errors that were present in their original form. In addition, portions of the data have been anonymized to protect the identity of the authors.

## FINDINGS

The fanfiction authors we interviewed believed that participating in fanfiction communities was both a source of emotional support and an avenue to improve the quality of their writing. They attributed these benefits to the varied forms of informal mentoring they received in the communities to which they belonged. Our participant observations led us to identify seven distinct attributes associated with the mentoring processes that participants described.

### Learning to Write, Learning about Life

Our findings provide evidence of the learning and mentoring occurring in fanfiction communities. It was striking, given how many online communities are not supportive [23], that every author we interviewed emphasized that support from the community contributed to their development as writers.

> *Writing fanfiction and getting instant feedback over the past couple of years has improved my writing significantly. I still have a long way to go, but I mostly avoid the standard pitfalls that young writers fall into such as poor flow and sentence construction, flat dialogue, and trivial descriptions.*
> - Author 19 (My Little Pony)

For most of the authors we interviewed, fanfiction was their first foray into creative writing, and it was fanfiction that taught them that they liked writing and gave them the desire to pursue original fiction. Author 11 has already produced an original fiction and attributed its existence to the confidence gained from writing fanfiction:

> *My original novel is due for publication next month. It's a fantasy fiction, part one of a planned four series. If it wasn't for the confidence I got from writing my fanfictions, I would never have even dreamed of self-publishing my own work.*
> - Author 11 (Harry Potter)

Authors also learned life lessons beyond writing from participating in fanfiction communities. Author 26 (My Little Pony) felt that the creative thinking involved in writing fanfiction helped him become more open-minded about other things, stating, "*If anything, thinking of premises for stories that can fit inside of canon has taught me to think outside of the box and expand things.*" Author 27 (My Little Pony) noted that in addition to learning about grammar, he has also learned to be more tolerant and willing to help others because of his community participation.

Some authors found emotional support or release from fanfiction. Author 16 felt that the response she received from the fanfiction community helped her get through a difficult period in her life:

> *I've received literally thousands of positive reviews and some truly wonderful letters and messages from people who have been genuinely touched by my writing, and it's been a massive confidence boost that helped me get through university without quitting and still helps me today if I'm feeling down.*
> - Author 16 (Harry Potter)

### Diverse Forms of Mentoring

In seeking to determine how the learning in fanfiction communities takes place, we uncovered diverse forms of mentoring. Authors we interviewed revealed that they experienced mentoring in contexts beyond a traditional dyadic mentor/mentee relationship. Fanfiction communities connect thousands of authors together and provide the opportunity for immediate and diverse feedback. Authors described a rich network of resources that they could draw on which included fanfiction reviews, FIMFiction.net user groups, beta readers, co-writers, Skype groups, podcasts, Internet Relay Chat (IRC), and other message boards and communities. Authors received and provided assistance through these channels in one-to-one, one-to-many, and many-to-one mentoring instances. Individually, these interactions with readers via story reviews or other authors via message boards may be too small to be considered mentoring, but in the aggregate they form a richer, more complete form of mentoring that is distributed across many participants and communication channels.

Differences in the affordances of fanfiction platforms give rise to distinct mentoring forms. For instance, FIMFiction.net includes a number of user groups, which several of the authors we interviewed said they relied on as a writing resource. Each user group has an associated forum, and in the writing help groups we observed, the members used these forums as a place to elicit advice on problematic aspects of their fanfiction stories (e.g. characterization and writing style) and to discuss general writing tips. The forums are publicly accessible, so any group member can respond to any post. In some cases, respondents would provide advice to the original poster independently without referencing what other respondents stated, but in other cases respondents would debate, correct, or agree with the advice provided by other respondents, often adding to it. For example, one poster described a rubric for creating an antagonist, and 21 group members posted completed rubrics while four members contributed additional information on the topic and commented on others' rubrics. One respondent added a 'map of evil' with different categories and characteristics of evil traits and left a caveat stating that "*not all antagonists have to be evil. In fact, making your reader sympathize with an antagonist is an advanced narrative technique.*" Another respondent who had previously posted a completed rubric replied "*Funny you should mention this. I plan to try this out with my next antagonist.*" These lively forum posts provide the original posters with a rich set of differing perspectives, something they would not have had access to if they had asked questions through more private, one-to-one channels to individual mentors (we term this *aggregation*, detailed below).

Many authors experienced mentoring in fanfiction communities as a reciprocal process. Author 6 described how the mentoring she received early on inspired her to provide the same support in a one-on-one mentoring context for a new writer:

> *I'll just add to the mentoring point - it's sort of come full cycle for me. When the girl PM'd [private messaged] me asking for advice, I did realise that I used to be her. Back in the day I wrote so badly that people flaming and trolling me would've been perfectly viable. Luckily I had people to push me up and advise me to turn me into the author I am today, so I found it really important to do exactly the same for her.*
> - Author 6 (Doctor Who)

The advice that Author 6 received in lieu of inflammatory comments also illustrates the supportive nature of the fanfiction community, where groups of authors are willing to provide feedback in order to help new writers grow (the influence of *affect* is described below).

Author 19 (My Little Pony) went so far as to express that one of the key reasons for participating in fanfiction was the review structure, stating "*I write fanfiction because you can get immediate feedback from an active fandom. I have to play my part as well, and give back to the community.*"

**The Importance of Reviews**
Fanfiction reviews present a particularly fertile ground for mentoring, where authors can learn from the tens, hundreds, or even thousands of reviews left on their work. Fanfiction.net's user base is much larger and more diverse than FIMFiction.net's, which may explain why it lacks the user group activity and community feeling of FIMFiction.net. However, stories posted on Fanfiction.net tend to receive a large number of reader reviews; therefore, we investigated the mentoring present in these reviews as another space for a distributed form of mentoring.

We observed that many reviewers provide thoughtful, detailed recommendations. These reviews let authors know specifically what they are doing well and what they need to work on. Author 23 recalled his experience as a beginning author as motivation to provide feedback:

> *I usually search for beginning authors and leave tips on how to improve. Such as better pacing, less telling, and most of all, trying to tone down a Mary Sue [self-insertion] type character or red and black alicorn OC [original characters]. What prompts me to leave feedback is that I used to be as inexperienced or beginning as a newbie on the site just like them not too long ago, so if I can help in any way, I'm going to try my hardest.*
> - Author 23 (My Little Pony)

Often, authors of substantive reviews identified themselves as part of the larger fandom community by demonstrating deep knowledge of the canon or fanon material and by providing corrective feedback for authors who made mistakes with regard to the canon:

> *Okay, I have one slight issue. Hermione doesn't learn the Patronus charm until her fifth year, so did being inside of Harry's head give her the ability to do it? If so, how would she know it was an otter if she wasn't supposed to have learned it yet.*

While these types of detailed reviews may be the most relevant for helping writers improve their craft, the cumulative encouragement contained in a collection of many shallow positive reviews (e.g. "Great story!") must not be discounted. We learned from participant observation and interviews that reviews are a crucial source of motivation for authors, and receiving a large number of any type of encouraging reviews reportedly helps authors persevere in an activity that requires substantial effort but provides few rewards aside from personal satisfaction. The cumulative encouragement present in large numbers of reviews (what we term *abundance*, described in more detail below) may also help authors understand if they are on the right track in terms of story aspects like plot, pacing, and writing style. Author 4 indicated she has written hundreds of reviews because she knows how important they are:

> *According to my General User Stats page, I've submitted 796 reviews on FFN dot net. I usually leave reviews for stories I really enjoy, or for stories I find don't have many reviews. I make it an effort to review everything I read. I know how much reviews are appreciated on this site.*
> - Author 4 (Doctor Who)

Overall, we found that reviewers were primarily providing supportive comments on the text rather than negative responses. Our interviews and observations suggest that shared interest in the fandom generates a sense of kinship and contributes to the positive atmosphere of the community. In the relatively infrequent instances when negative comments were posted, we observed far more community members come to the defense of the author of the story being criticized. One controversial fanfiction story contained several flame reviews but even more reviews like this one defending the author:

*Just ignore the stupid tart-lover [reference to a Doctor Who character] who is leaving those horrid reviews. She can take all her stuffy little remarks and shove them where the light don't shine. This is a fantastic story, I just had a shitty day and this story has really helped me smile.*

This quote also illustrates the fact that many fanfiction reviewers are engaged in dialogue with other reviewers and authors and are considering other feedback and notes as they construct their reviews. In other words, the mentoring activity is not focused exclusively on a single author, but demonstrates high levels of interaction among reviewers, what we term *accretion* and *acceleration* (described below). This emergent, complex structure within the mentoring activity is a key attribute of distributed mentoring.

**The Seven Attributes of Distributed Mentoring**

The themes emerging from the author interviews guided our participant observations and led us to identify seven attributes that describe the mentoring processes we observed. These attributes are: *aggregation, accretion, acceleration, abundance, availability, asynchronicity,* and *affect*. Despite some differences in mentoring style across platforms (e.g. user groups on FIMFiction.net), we found evidence of all seven attributes across the fandoms and platforms we studied.

*Aggregation*: Authors experienced mentoring from a number of different sources. Powered by the openly networked nature of computer-mediated communication, the authors in our study received numerous small pieces of advice in the form of story reviews or discussion forum responses that, as in distributed cognition [30], created a whole far greater than the sum of its parts. The feedback provided independently by each participant may not be sufficient in either depth or longevity to be considered mentoring on its own. Instead, authors are being mentored in the aggregate by the fanfiction community at large.

> *I can't think of any one person [in particular who has influenced me], I take inspiration from a number of people.*
> - (Author 19, My Little Pony)

Based on our observation of reviews, an author who receives even ten reviews on his or her work is likely to receive five targeted pieces of feedback in addition to several other encouraging remarks. Together, this aggregated feedback may let authors know what they have done well and what is lacking in their fanfiction stories. Even numerous shallow positive reviews can show authors collectively that they are on the right track.

*Accretion*: Reviewers themselves interacted with each other in a persistent, cumulative fashion, enabling an accretion of knowledge to facilitate the author's learning process. The mentoring activity pertaining to the creation of an antagonist is an example of this (described on p. 6), where the process of developing a compelling antagonist in one's story was carefully deconstructed by a group whose members built upon and extended previously posted commentary. Through these distributed postings and exchanges, the information accreted over time, enabled by the persistent nature of the forum. The advice that authors received from this mentoring experience included targeted, cumulatively sophisticated and detailed feedback on their fanfiction stories and thorough answers to their questions posted on forums or in authors' notes.

*Acceleration*: It is sometimes the case that authors receive conflicting feedback in the variety of reviews of their work. While this may at first seem confusing, we found that discussions and conflicts between reviewers typically yield a more complex and nuanced body of feedback as reviewers point out holes in each other's arguments or reinforce each other's statements through agreement, cite deep fandom knowledge, and accelerate the process of learning through active discussion. We observed an interwoven network of reviewer responses delving into highly specialized knowledge and buttressing subtle points of argument.

*Abundance*: The sheer volume of review responses contributes to the nature of distributed mentoring. By itself, a shallow positive comment is relatively meaningless. But if an author receives dozens or hundreds of such comments, the mere presence of positive feedback in such quantities provides important guidance to the writer. Additionally, the existence of multiple channels of feedback and mentoring, such as private messaging, author's notes, reviews, Skype chats, etc. contribute to the multiplicity and abundance of distributed mentoring.

> *If I've got an idea, but need to spin it out a bit more, I'll just pop into one of my 30+ member Skype chats and toss it out there. I can always expect good feedback there.*
> - (Author 22, My Little Pony)

*Availability*: On Fanfiction.net and FIMFiction.net, the interactions of individual participants are visible to the larger community. The persistent and public nature of online text-based communication ensures that reviews are available not just immediately after a story is written, but indefinitely, months or years into the future. This availability in turn facilitates long-term exchanges and relationships between reviewers and authors, as authors can post their stories and learn from their reviewers (and reviewers can learn from each other) years into the future, thus further broadening the effect of distributed mentoring.

*Asynchronicity*: The public, durable text interactions allow for asynchronous contributions by any person, at any time, no matter the location, enabling reviewers from multiple time zones around the globe to easily and continuously view and reply to other reviews. Authors might post a story, go to sleep, and in the morning wake up to multiple thoughtful and lengthy critiques of their work, which they or other reviewers can then respond to as they wake up in their own time zones. This asynchronicity adds depth, immediacy, and continuity to a mentoring conversation. The power of asynchronous communication, especially in international communities operating in different time zones and outside work or school hours, has been noted in previous research on online community [5]. It is a crucial enabling factor for distributed mentoring.

*Affect*: In interviews, on forums, and in their author's notes, fanfiction writers repeatedly mentioned positive affect or emotion generated by the encouragement and inspiration they received from reviews and feedback (see "The Importance of Reviews" above). Negative reviews or flames were moderated by other reviewers creating a supportive community response. It is important not to ignore the emotional and supportive aspects of mentoring, especially with respect to adolescents who are forming their identities and building self-efficacy. The act of writing, even nonfiction writing, can be an emotionally charged event where otherwise highly skilled individuals find themselves unable to produce the written communication that is crucial for success in almost every field of endeavor. Authors in these communities cited the emotional support they received as key to their growth as writers.

These attributes and tendencies themselves are interwoven, as each contributes to and facilitates other attributes, further contributing to the interdependent and complex nature of distributed mentoring. For example, participants must be able to view the responses of others (*availability*) so they can take these diverse perspectives into consideration when creating their own responses or reviews (*accretion*), and respondents can directly contradict or support other respondents (*acceleration*). As the fanfiction stories we sampled increased in their number of reviews (*aggregation*), we saw an increase in the amount of discussion present in these reviews, indicating that interactions in distributed mentoring become more complex and richer with increasing numbers of participants and responses.

Any phenomenon that draws millions of young people to participate and generate literary content, especially in a generation criticized for their inability to write coherently [11,23,59,62], is deserving of closer attention. In particular, understanding in more depth how networked publics facilitate this new type of learning and mentoring is vital for the future of formal and informal education as more and more young people spend a significant fraction of their waking hours online [23]. The seven attributes we identified form the beginning of an understanding of the underlying structure of this new kind of mentoring.

## DISCUSSION

We set out to understand what adolescents and young adults were learning from their participation in online fanfiction communities, how this learning took place, and what part mentoring played. We initially anticipated that we would uncover traditional mentoring relationships where less experienced writers learned from more experienced writers; instead, during our investigation, we encountered a rich and interwoven tapestry of interactive, cumulatively sophisticated advice and informal instruction that added up to a networked experience we have termed distributed mentoring.

Grounded in Hutchins' [30] theory of distributed cognition, distributed mentoring takes as its base the fact that knowledge is distributed across people and artifacts and the aggregation of this knowledge is greater and qualitatively different from the knowledge residing in any one individual or artifact. As a result, distributed mentoring is most likely to be found in online affinity spaces [24,25]. Consistent with Gee's description of an affinity space but with a particular focus on mentoring processes, we found that mentoring resources were shared socially to extend and broaden the influence and knowledge from the domain of the individual to the group, so that a greater effect was achieved than through traditional mentoring. This was made possible and facilitated specifically by networked publics and thus did not occur in fan communities before the advent of widespread online communication. As in distributed cognition, mentoring processes are distributed among humans and artifacts in the environment, and mentoring systems are formed where complex structure emerges in ways that enhance the reach of any single individual. However, distributed mentoring extends Hutchins' concept of distributed cognition by focusing specifically on mentoring processes as well as the influence of networked technologies' distinct affordances.

Distributed mentoring also pushes at the framework for traditional mentoring models established by Dawson [18]. Most of these design elements become irrelevant or extremely varied within the informal structure of the online fanfiction community. Selection and matching of mentoring relationships, crucial components of most traditional models, are entirely self-driven in fanfiction communities. Roles, cardinality (number of relationships), ties, relative seniority, and time vary both by individuals and by task (e.g. proofreading versus plotting). There is no mentorship training, the rewards are all intrinsic, and the monitoring is crowdsourced. Yet, powerful and effective mentorship occurs daily within these sites for thousands of authors. Therefore, a new mentoring model is happening that is uniquely supported by the affordances of networked technologies and for which we must account. The seven attributes we identified—*aggregation, accretion, acceleration, abundance, availability, asynchronicity,* and *affect*—illustrate how the affordances of networked technologies give rise to distributed mentoring and set it apart from traditional mentoring.

Previous work has detailed the distinct affordances of networked technologies [10,61] as well as how they support new social forms, such as participatory culture [35,36] and connected learning [33]. The attributes of distributed mentoring we have identified are consistent with the characteristics of participatory culture (e.g. low barriers to entry, informal mentorship) and supported by the core properties of connected learning (shared purpose, production-centered, openly networked). However, the theory of distributed mentoring extends prior work by focusing explicitly on the interplay between mentoring and networked technologies, showing how the affordances of networked technologies shape mentoring processes in new ways.

For instance, the persistent quality of computer-mediated communication (CMC) makes it possible for authors to consider multiple voices together, both in the moment and at a later date, supporting the attributes of *aggregation, accretion, acceleration*, and *asynchronicity* [19,53]. This may be challenging in non-CMC contexts, where human memory limits the number of previous responses a participant can consider. In addition, the public nature of communicating in networked environments like fanfiction sites allows for a high level of interaction among reviewers, who can react to and build on each other's comments asynchronously and over the long term. This quality of networked communication supports the attributes of *abundance, availability*, and *affect*. In contrast, traditional ideas of mentoring are grounded primarily in one-to-one relationships, which are easier to maintain in the physical world. Mentoring facilitated by the internet can move past the limitations of the physical world to allow numerous mentors to engage each other on the same topic asynchronously when providing advice to mentees.

**Benefits of Distributed Mentoring**
The fanfiction authors emphasized how much they valued the mentoring that they received from the larger community, which they viewed as an important part of their growth as writers as well as a morale boost to keep writing (*affect*). The unique style of distributed mentoring available in fanfiction communities is one reason why young authors often choose to practice writing by creating fanfiction before attempting to write original fiction. Authors mentioned that the feedback that they received from the fanfiction community was a key reason for their participation and that they viewed fanfiction as an excellent training ground for aspiring writers who hope to produce original fiction one day. One published writer who had started out writing fanfiction commented in a blog post that she received more feedback in a week on her fanfiction than she had in several years on her original fiction. In the communities that we investigated, participants were very supportive of each other and eager to provide informal mentoring. The communities also provided a knowledge base and instant connection based on the fandom for anyone wishing to participate.

The benefits derived from distributed mentoring are similar to the benefits associated with distributed cognition [30]. Distributed mentoring provides artifacts that are embodied outside the author, such as forum responses, reviews, private messages, and author's notes. These artifacts, as in distributed cognition, allow for the coordination and aggregation of mentoring efforts. As well, the distribution of access to information among these artifacts in the contexts we studied allowed for error checking in the form of agreement, disagreement, and discussion among respondents, just like in a cognitive system.

The whole of the mentoring experience provided by distributed mentoring is greater than the sum of its parts. Distributed mentoring creates richer mentoring experiences than individual mentors alone could provide, much like how distributed cognition allows cognitive systems to accomplish tasks that would be otherwise impossible for individuals to complete. A single mentor can provide only one perspective, and if a writer asks multiple mentors independently, it will be up to the author to decide which pieces of advice are sound. In FIMFiction.net group discussions, authors can learn from the different pieces of advice individually and also learn from the discussion between the respondents. In these discussions, bad advice is corrected, good advice is validated, and more nuanced answers to questions form as a result of the respondents building off of each other (*aggregation*, *accretion*, and *acceleration*). On Fanfiction.net, authors are able to learn from reviews that focus on different aspects of the story and from reviewers who may be more critical or more positive. In addition, inflammatory reviews are moderated by reviewers' responses, and responses to author's notes provide targeted assistance for authors.

As in other affinity spaces, participants in online fanfiction communities can be any age, from any location, and of any experience level. There are no defined roles of "master" and "apprentice" in these communities because one author might possess deep knowledge on a certain topic, e.g. characterization, and yet be lacking in another aspect, e.g. grammar. This diversity of expertise and the lack of defined roles in online fanfiction communities are of benefit in distributed mentoring situations because different participants can bring up different perspectives in their feedback. Also, because distributed mentoring is peer-based, or horizontal, rather than hierarchical, an author who asks for assistance can turn around and provide feedback for another author on a different topic. The authors we interviewed revealed the important role that reciprocity played in the fanfiction community, indicating that the help they received inspired them to help others (*acceleration*).

Participation in affinity spaces and participatory cultures is predicated on participants' intersubjective understanding of the context in which they are taking part, just like the ship navigators and pilots described by Hutchins [30] and Hutchins and Klausen [31]. In our observations, the substantial number of reviews involving some form of fandom discussion provided evidence that authors and reviewers share a common base of language and knowledge. Indeed, their shared knowledge goes beyond what is represented in the canonical literature to include fan canon notions of untold character backstories, favored relationship pairings, and popular plot conspiracies. Our analysis suggests that the communitarian ethos present in the civil, reciprocal exchanges in online fanfiction communities is likely encouraged by the membership that authors feel as participants in the community.

**Implications**
Previous research investigating fanfiction has looked at how the mentoring that authors receive via interactions with other community members impacts their identity [6–8] and improves their writing ability [35]. However, there was a lack of research into the relationship between these individual-level effects and the group-level qualities of fanfiction communities, such as the massive scale of mentoring; its distributed nature across individuals, time, and platforms; and the distinct characteristics of author-mentor and mentor-mentor interactions. In particular, our work extends Black's research [6–8] by broadening the focus beyond a small number of fanfiction authors, a single fanfiction repository (Fanfiction.net), and a single fanfiction genre (anime). This broader view allowed us to detect common patterns of mentoring across different fanfiction communities, indicating that distributed mentoring is not limited to one specific context or group of participants. In addition, by focusing on mentoring at the group- rather than the individual-level, we were able to document the patterns and effects of multiple sources of feedback in the aggregate. Central to this analysis was an enumeration of the distinct affordances of networked technologies and how they distinguish distributed mentoring from traditional models of mentoring. By extending Black's research [6–8] in this way, our work provides new insight into leveraging the affordances of networked publics to support and promote collaborative learning.

For instance, the insights from the present study suggest that distributed mentoring could be successfully applied to formal educational contexts as a useful peer mentoring mechanism, building on previous related work exploring the incorporation of fanfiction in classroom settings [43,49]. An internet-based peer review system joining a number of classrooms together would help facilitate distributed mentoring. Students in one classroom could be tasked with contributing comments to the work of students in other classrooms. By interacting with students from another classroom, located in a different school or perhaps even a different city or country, the pseudonymous nature of distributed mentoring would be maintained. This geographic and emotional distance could help to prevent feelings of inhibition that students might feel if asked to give and receive feedback from the classmates they see and interact with on a daily basis. Moreover, students would be able to benefit from encountering, considering, and responding to perspectives different from their own [29,52]. Gee states that "[networked publics] are allowing people with no official credentials from formal institutions to become experts," and this development of expertise can be encouraged in classroom settings [25]. Our findings suggest that students should be encouraged to leave a small amount of feedback for each piece of writing they are to review, making sure to read and build upon previous comments. In this way, students could learn from the interactions of their reviewers in addition to the different perspectives presented in each review, which is what makes distributed mentoring such a rich learning experience. The attributes observed in spontaneous forms of distributed mentoring could be used to evaluate and refine these planned mentoring systems. Black [8] saw adolescents motivated to participate in fanfiction spaces by the experience of procedural knowledge (as contrasted to propositional knowledge) and opportunities to contribute to the function of the space. Therefore, a distributing mentoring model in the classroom would need to maintain such features to sustain student engagement.

In addition, distributed mentoring holds implications for the design of online affinity spaces. As the differences between Fanfiction.net and FIMFiction.net illustrate, website design facilitates or discourages different types of interactions. The common features across these sites, such as comment threads and discussion groups, suggest that these interest-driven communities already afford rich distributed mentoring. The power of distributed mentoring comes in part from the numerous responses or reviews that authors receive. As an author receives more replies, the mentoring experience becomes richer. More could be done, however, to help authors who are not receiving a lot of feedback to benefit from distributed mentoring. Although it is currently possible to browse fanfiction sites by number of reviews, readers have to intentionally seek out stories with few reviews. A simple recommender that directs community members who have just left a review to similar stories that have few or no reviews may help to channel feedback to stories that need it most. Such a system could also steer readers and writers to similar stories, thus affording more community formation and participation. Affective features such as easy ways to "like" or categorize stories could enhance positive experiences. Developers could also enhance the mentoring experience for contributors who receive large numbers of reviews by adding tools that use text mining techniques to automatically synthesize common themes or contrasting viewpoints.

**LIMITATIONS AND FUTURE DIRECTIONS**

As is inherent in most qualitative research, our site for this study was limited. We investigated only three fandoms and two fanfiction repositories in detail; therefore, our findings may not generalize to mentoring in online affinity groups at large. There may be unique aspects specific to these two fanfiction communities that encourage members to leave more feedback. The communities we observed were very civil, as shown by the low amount of trolling present in story reviews. This civility was surprising because readers can leave anonymous reviews on Fanfiction.net, and anonymous forms of online communication have a reputation for being susceptible to trolling [46]. This civility could in part be a legacy of the benign atmosphere found in early fanfiction zines' Letters of Comment; it may be that other affinity spaces that lack this legacy may not be supportive enough to sustain distributed mentoring [34].

We are not able to claim for certain that distributed mentoring directly affects writing ability. We set out to understand and describe the type of mentoring present in online fanfiction communities, leaving the verification of its effect to later research. Many of the authors we interviewed believed that the feedback they received from the fanfiction community helped improve their writing, but our research was observational in nature, not experimental. There are a number of other potential learning resources in these young authors' lives that have certainly impacted their writing, although we think on the basis of our findings that distributed mentoring is one of them.

As distributed mentoring is a nascent concept, there are many opportunities to explore regarding its effectiveness and application, as well as its limitations. More work is needed to determine how distributed mentoring impacts writing quality, especially in comparison with other types of mentoring or formal education. Does the computer-mediated nature of distributed mentoring allow for a more diverse and therefore richer mentoring experience compared to in-person peer mentoring? Does writing ability improve more when writers experience distributed mentoring compared to communicating independently with multiple mentors via the internet?

Additionally, we did not investigate how website structure and design aids or impedes distributed mentoring, although we did observe different styles of distributed mentoring on different websites. Writers on FIMFiction.net frequently received help on user group forums, while writers on Fanfiction.net often received help from reviews in reply to authors' notes. We wonder if a particular style of feedback mechanism (e.g., forums) yields superior mentoring experiences. Also, do more opportunities for users to leave feedback actually yield more feedback, and what is the quality of this feedback?

## CONCLUSION

Online fanfiction communities represent affinity spaces in which authors, brought together by a passion for the fandom, provide support for each other in their creative endeavors. Through in-depth interviews and participant observations, we identified and documented a unique style of mentoring occurring in these communities that we named *distributed mentoring*, inspired by Hutchins' [30] theory of distributed cognition. In our investigation, we observed and described seven key attributes of distributed mentoring that distinguish it from traditional, offline forms of mentoring and tie it to the particular affordances of networked publics: *aggregation, accretion, acceleration, abundance, availability, asynchronicity,* and *affect*. Distributed mentoring involves numerous participants answering questions, leaving feedback, and engaging each other in asynchronous conversations that in the aggregate yield a positive mentoring experience. The authors we interviewed valued the feedback they received from the community and viewed it as an important part of their growth as writers. Aside from its role in online communities, distributed mentoring may also present opportunities, as described in the implications section, for collaborative learning in other learning contexts, including formal educational contexts.

## ACKNOWLEDGMENTS

We thank the reviewers who provided thoughtful and detailed comments on previous iterations of this document. The final paper has been greatly improved by their feedback. We also extend gratitude to our participants in this study and our colleague, Jeff Giorgi.

## REFERENCES

1. Alexa. 2015. Site Overview – Fanfiction.net. Retrieved May 4, 2015 from http://www.alexa.com/siteinfo/fanfiction.net
2. Alexa. 2015. Site Overview – Fimfiction.net. Retrieved May 4, 2015 from http://www.alexa.com/siteinfo/fimfiction.net
3. Ayşegül Kuglin Altintaş. 2013. A new Hermione: re-creations of the female Harry Potter protagonist in fan fiction. *Z Anglist Am* 61, 2: 155–173.
4. Cecilia R. Aragon and Alison Williams. 2011. Collaborative creativity: a complex systems model with distributed affect. In *Proceedings of the SIGCHI Conference on Human Factors in Computing Systems (CHI '11)*, ACM, 1875–1884.
5. Cecilia R Aragon, Sarah Poon, Andres Monroy-Hernandez, and Diana Aragon. 2009. A tale of two online communities: Fostering collaboration and creativity in scientists and children. *Creativity & Children (C&C '09)*, ACM, 9–18.
6. Rebecca W. Black. 2006. Language, culture, and identity in online fanfiction. *E-Learning* 3, 2: 15.
7. Rebecca Black. 2007. Digital design: English language learners and reader reviews in online fiction. In *A New Literacies Sampler*, Michele Knobel and Colin Lankshear (eds.). Peter Lang, New York, NY, 115-136.
8. Rebecca W Black. 2008. *Adolescents and Online Fanfiction*. Peter Lang, New York, NY.
9. Tom Boellstorff, Bonnie Nardi, Celia Pearce, and T.L. Taylor. 2012. *Ethnography and Virtual Worlds: A Handbook of Method*. Princeton : Princeton University Press, Princeton.
10. Danah Boyd. 2007. The role of networked publics in teenage social life. In *Youth, Identity, and Digital Media*. MIT Press, Cambridge, MA, 119–142.
11. Tom Bradshaw. 2004. *Reading at Risk : A Survey of Literary Reading in America*. Washington, D.C. : National Endowment for the Arts, Washington, D.C.
12. Kelly Chandler-Olcott and Donna Mahar. 2003. Adolescents' anime-inspired "fanfictions": An exploration of multiliteracies. *J of Adoles & Adult Lit* 46, 7: 556–566.
13. Kathy Charmaz. 2006. *Constructing Grounded Theory: A Practical Guide Through Qualitative Analysis*. Sage, London.
14. Michelle Chatelain. 2012. Harry Potter and the prisoner of copyright law: Fan fiction, derivative works, and the fair use doctrine. *Tulane J of Tech & Intellect Pro* 15: 199–319.
15. Mark Chen. 2012. *Leet Noobs: The Life and Death of an Expert Player Group in World of Warcraft*. New York : Peter Lang, New York.
16. Computing Research Association. 2005. CRA-W Canadian Distributed Mentor Project. Retrieved July 27, 2015 from http://archive2.cra.org/Activities/craw_archive/cdmp/index.php
17. Katie Davis. 2010. Coming of age online: The developmental underpinnings of girls' blogs. *J of Adolesc Res* 25, 1: 145–171.
18. Phillip Dawson. 2014. Beyond a definition: Toward a framework for designing and specifying mentoring models. *Ed Res* 43, 3: 137–145.

19. Pierre Dillenbourg. 2005. Designing biases that augment socio-cognitive interactions. In *Barriers and Biases in Computer-Mediated Knowledge Communication: And How They May Be Overcome*, Rainer Bromme, Friedrich W Hesse and Hans Spada (eds.). Kluwer Academic Publisher, Dordrecht, Netherlands, 243–264.
20. Elizabeth Dutro and Monette C. McIver. 2011. Imagining a writer's life: Extending the connection between readers and books. In *Handbook of Research on Children's and Young Adult Literature*, Shelby Anne Wolf, Karen Coats, Patricia Enciso and Christine A. Jenkins (eds.). Routledge, New York, 92–107.
21. Ellen A Ensher, Christian Heun, and Anita Blanchard. 2003. Online mentoring and computer-mediated communication: New directions in research. *J of Voc Behav* 63, 2: 264–288.
22. FIMFiction.net. 2015. FIMFiction.net Statistics. Retrieved July 19, 2015 from http://www.fimfiction.net/statistics
23. Howard Gardner and Katie Davis. 2013. *The App Generation: How Today's Youth Navigate Identity, Intimacy, and Imagination in a Digital World*. Yale University Press, New Haven
24. James Paul Gee. 2004. *Situated Language and Learning: a Critique of Traditional Schooling*. New York : Routledge, New York.
25. James Paul Gee. 2013. *The Anti-Education Era: Creating Smarter Students Through Digital Learning*. New York City : Palgrave Macmillan.
26. Darlene Rose Hampton. 2015. Bound princes and monogamy warnings: Harry Potter, slash, and queer performance in LiveJournal communities. *Transform Works & Cult* 18.
27. Ashley Hinck. 2012. Theorizing a public engagement keystone: Seeing fandom's integral connection to civic engagement through the case of the Harry Potter Alliance. *Transform Works & Cult* 10.
28. James Hollan, Edwin Hutchins, and David Kirsh. 2000. Distributed cognition: Toward a new foundation for human-computer interaction research. *ACM TOCHI* 7, 2: 174–196. http://doi.org/10.1145/353485.353487
29. Sylvia Hurtado, Mark Engberg, Luis Ponjuan, and Lisa Landreman. 2002. Students' precollege preparation for participation in a diverse democracy. *Res in Higher Ed* 43, 2: 163–186.
30. Edwin Hutchins. 1995. *Cognition in the Wild*. MIT Press, Cambridge, Mass.
31. Edwin Hutchins and Tove Klausen. 1996. Distributed cognition in an airline cockpit. In *Cognition and Communication at Work*, Yrjö Engeström and David Middleton (eds.). Cambridge University Press, 15–34.
32. Mizuko Itō. 2010. *Hanging Out, Messing Around, and Geeking Out: Kids Living and Learning with New Media*. MIT Press, Cambridge, Mass.
33. Mizuko Itō, Kris Gutiérrez, Sonia Livingstone, Bill Penuel, Jean Rhodes, Katie Salen, Juliet Schor, Julian Sefton-Green, and S. Craig Watkins. 2012. *Connected Learning: An Agenda for Research and Design*. Digital Media and Learning Research Hub.
34. Henry Jenkins. 1992. *Textual Poachers: Television Fans & Participatory Culture*. Routledge, New York.
35. Henry Jenkins. 2006. *Convergence Culture: Where Old and New Media Collide*. New York University Press, NY.
36. Henry Jenkins. 2009. *Confronting the Challenges of Participatory Culture: Media Education for the 21st Century*. The MIT Press, Cambridge, MA.
37. Henry Jenkins. 2012. "Cultural acupuncture": Fan activism and the Harry Potter Alliance. *Transform Works & Cult* 10.
38. Henry Jenkins and Kurt Squire. 2011. *Video Games and Learning: Teaching and Participatory Culture in the Digital Age*. New York : Teachers College Press, NY.
39. Jessica Elizabeth Johnston. 2015. Doctor Who–themed weddings and the performance of fandom. *Transform Works & Cult* 18.
40. Debi Khasnabis, Catherine H. Reischl, Melissa Stull, and Timothy Boerst. 2013. Distributed mentoring: Designing contexts for collective support of teacher learning. *Eng J* 102, 3: 71–77.
41. Colin Lankshear and Michele Knobel. 2007. Researching new literacies: Web 2.0 practices and insider perspectives. *E-Learning* 4, 3: 224–240.
42. Jean Lave and Etienne Wenger. 1991. *Situated Learning: Legitimate Peripheral Participation*. Cambridge University Press, Cambridge.
43. Margaret Mackey and Jill Kedersha Mcclay. 2008. Pirates and poachers: Fan fiction and the conventions of reading and writing. *Eng in Ed* 42, 2: 131–147.
44. Kerri Mathew and Devon Adams. 2009. I love your book, but I love my version more: Fanfiction in the English language arts classroom. *ALAN Review* 36, 3: 35-41.
45. Sharan B. Merriam. 2014. *Qualitative Research: A Guide to Design and Implementation*. Wiley, Hoboken.
46. Michael J. Moore, Tadashi Nakano, Akihiro Enomoto, and Tatsuya Suda. 2012. Anonymity and roles associated with aggressive posts in an online forum. *Comp in Hum Behav* 28, 3: 861–867.
47. Mollie E. Nolan. 2006. Search for original expression: Fan fiction and the fair use defense. *SIU Law J* 30, 3: 533-571.
48. Nick Pidgeon and Karen Henwood. 1996. Grounded theory: Practical implementation. In *Handbook of Qualitative Research and Methods for Psychology and the Social Sciences*, J. T. E. Richardson (ed.). Wiley, Oxford.
49. Kevin Roozen. 2009. "Fan Fic-ing" English studies : A case study exploring the interplay of vernacular literacies and disciplinary engagement. *Res in Teach of Eng* 44, 2: 136–169.

50. Jim Sandherr, Asia Roberson, Caitlin K. Martin, Ugochi Acholonu, and Denise Nacu. 2014. Distributed mentorship: Increasing and diversifying youth access to learning networks. *Panel at the 6th annual Digital Media and Learning Conference (DML '14)*.
51. Aaron Schwabach. 2009. The Harry Potter lexicon and the world of fandom: Fan fiction, outsider works, and copyright. *U of Pitt Law Rev* 70, 3: 387–434.
52. Robert L. Selman. 1975. Level of social perspective taking and the development of empathy in children: Speculations from a social-cognitive viewpoint. *J of Moral Ed*.
53. Gerry Stahl, Timothy Koschmann, and Daniel D. Suthers. 2006. Computer-supported collaborative learning. In *The Cambridge Handbook of the Learning Sciences*, R. Keith Sawyer (ed.). Cambridge University Press, Cambridge.
54. Constance A. Steinkuehler. 2004. Learning in massively multiplayer online games. In *Proceedings of the 6th International Conference on Learning Sciences (ICLS '04)*, 521–528.
55. Leanne Stendell. 2005. Fanfic and fan fact: How current copyright law ignores the reality of copyright owner and consumer interests in fan fiction. *SMU Law Rev* 58: 1551.
56. Anselm L. Strauss and Barney G. Glaser. 1967. *The Discovery of Grounded Theory : Strategies for Qualitative Research*. Aldine Pub. Co., Chicago.
57. Rachel L. Stroude. 2010. Complimentary creation: Protecting fan fiction as fair use. *Marquette Intellect Prop Law Rev* 14, 1: 191.
58. Catherine Tosenberger. 2008. Homosexuality at the online hogwarts: Harry Potter slash fanfiction. *Child Lit* 36, 1: 185–207.
59. Sherry Turkle. 2011. *Alone Together : Why We Expect More from Technology and Less from Each Other*. New York : Basic Books.
60. Kevin Veale. 2013. Capital, dialogue, and community engagement—My Little Pony: Friendship Is Magic understood as an alternate reality game. *Transform Works & Cult* 14.
61. Joseph B. Walther. 1996. Computer-mediated communication: Impersonal, interpersonal, and hyperpersonal interaction. *Comm Res* 23, 1: 3–43.
62. Emily Weinstein, Zachary Clark, Dona Dibartolomeo, and Katie Davis. 2014. A decline in creativity? It depends on the domain. *Creativ Res J* 26, 2: 174–184.